\documentclass[11pt]{amsart}
\usepackage{geometry}                
\usepackage{graphicx}
\usepackage{amssymb}
\usepackage{epstopdf}
\usepackage{multicol}
\usepackage{float}
\usepackage[bottom]{footmisc}
\usepackage{url}
\title{Uncertainty in Security: \\managing cyber senescence}
\author{Martijn Dekker}
\date{November 28, 2025}                                           

\graphicspath{{Fig/}}

\begin{document}

\begin{abstract}
My main worry, and the core of my research, is that our cybersecurity ecosystem is slowly but surely aging and getting old and that aging is becoming an operational risk. This is happening not only because of growing complexity, but more importantly because of accumulation of controls and measures whose effectiveness are uncertain. I introduce a new term for this aging phenomenon: “cyber senescence”.
I will begin my lecture with a short historical overview in which I sketch a development over time that led to this worry for the future of cybersecurity. It is this worry that determined my research agenda and its central theme of the role of uncertainty in cybersecurity. My worry is that waste is accumulating in cyberspace. This waste consists of a multitude of overlapping controls whose risk reductions are uncertain. Unless we start pruning these control frameworks, this waste accumulation causes aging of cyberspace and could ultimately lead to a system collapse.
\end{abstract}

\maketitle 

\begin{multicols}{2} 
\section{Introduction}
In this lecture\footnote{This paper is a slightly edited and translated version of the author's inaugural lecture "onzekerheid in veiligheid" at the University of Amsterdam.} I will talk about uncertainty. With “uncertainty”, I do not mean that the future is unknown and likelihoods of events happening, but the fundamental lack of information that prevents one of even determining the probability distribution and impacts of possible events. This uncertainty is caused by vulnerable software, complexity of the system and the unfalsifiability of security claims. I will explain these points.

In the 1970’s, the predecessor of Internet emerged from the research programme ARPANET of the American ministry of defence. Originally, the network consisted of just a few computers connected with noisy telephone lines. Researcher Bob Thomas created the computer program “creeper” that moved across this network leaving the message “I’m the creeper, catch me if you can” on the computers. As a response, Ray Tomlinson wrote the program “reaper” that was able to chase Creeper and remove it from a system. The first computer virus and anti-virus software were born. A more extensive description can be found in The Pandora Tech Blog (\cite{reaper}).

In 1983, the movie “War Games” (cf. \cite{wargames}) came out. In this movie, a piece of software nearly caused a nuclear war. The movie came out in the middle of the cold war and cyber espionage developed quickly. It led the United States Department of Defense to publish in 1983 and updated in 1985, the so-called Orange Book (\cite{orange}), which later became ISO/IEC 15408 in 1999. This publication set out in precise terms which measures software developers were required to take to produce trustworthy software. 

In retrospect, it is remarkable to realise that such principles were formulated so long ago, especially given our widespread familiarity today with the many vulnerabilities and disruptions in the software we all use daily.

On June 19th, 2024, the American cybersecurity company Crowdstrike issued a small software update for their Falcon Sensor software. Falcon Sensor is software designed to protect computer systems against cyber-attacks. It can detect and respond to attacks. To be able to do that, the software needs to be installed on many systems and it requires high level authorisations and access.
Millions of computers in the world use this software to protect them and Crowdstrike issues updates and updated configurations with high frequency of sometimes more than one per day. They do this for good reasons: our digital world has become big and complex, and these updates share information and hence reduce uncertainty.

Unfortunately, the update of June 19th contained a mistake. One of the software engineers had made a mistake that was not discovered until the update was installed on millions of computers. The effect of this mistake was that about 8,5 million computers worldwide crashed and could not be restarted. As a result, about 60\% of the Fortune 500 and over 24,000 other customers of Crowdstrike faced disturbances and outages of their services.  Hotels, hospitals and harbours struggled with their operations, had to change their schedules or had to shut down completely. The impact of this incident was felt worldwide but some countries were not impacted at all: China, Russia and Iran do not want to use American software due to political reasons, and hence were not directly impacted.

The Crowdstrike incident is probably the largest IT incident worldwide in history (cf. \cite{crowdstrike}). But it certainly is not the only impactful incident. In October 2025, another disruption occurred at the company Cloudflare (cf. \cite{cloudflare}). Their bot-management system received a faulty update which led to massive unavailability of services worldwide.

These incidents illustrate several aspects of the world we live in and in which organisations are operating. It illustrates how the world has changed since the arrival of Creeper and Reaper. Note that both incidents, Crowdstrike and Cloudflare, were caused by misbehaving cybersecurity systems, that are supposed to protect computers. I think this is relevant and I will explain why.

\section{Security}
Cybersecurity frameworks have evolved over time. From the very beginning of computers and computer networks, scholars were very aware of the necessity of security. Security by design is not a new notion, although it is recently being reinvented. The earlier mentioned Orange Book of the American ministry of defence contains many measures software producers must implement in software before they can sell it. These measures range from preventing of unauthorised access to the ability to find traces of misuse by mandatory logging. 

Over the years, the collection of controls and measures has grown significantly. The most well-known framework for cybersecurity is the cybersecurity framework (CSF) of NIST, the American National Institute of Standards (\cite{nist}). The last version (2.0) of this framework describes about 1200 controls while the first version only contained 400 controls. Next to the huge and still growing number of controls, the NIST framework also grew by additions of new types of controls. I will give two examples.

Firstly, the framework not only contains controls that are designed to prevent security incidents to happen, but it also contains controls that are needed to be able to respond and recover once an incident does happen. This is an essential insight in modern cybersecurity: it is impossible to prevent all security incidents and organisations must prepare for incidents happening and be able to respond to an incident to minimise impacts. This is why we rather talk about cyber resilience instead of cyber security.

\begin{figure}[H]
\includegraphics[width=0.5\textwidth]{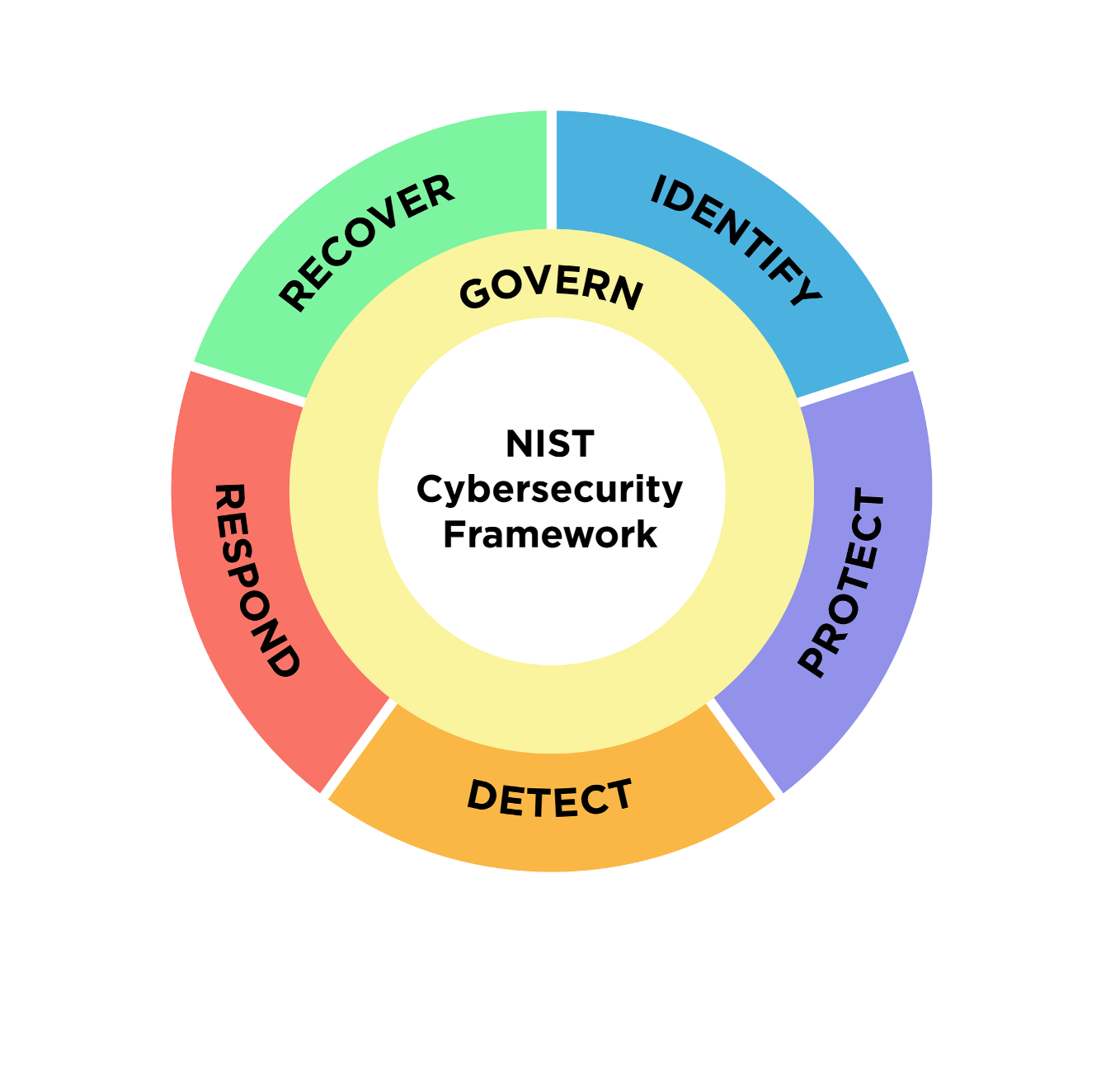}
\caption{NIST CSF 2.0}\label{fig1}
\end{figure}

Secondly, in the latest version of the NIST CSF a new category of controls has been added. This new category, called “governance”, contains controls needed to govern security (see figure \ref{fig1}). This category was added because cybersecurity is no longer a purely technical topic, dealt with by computer scientists and programmers. Instead, it has become a topic that requires steering, management and decision making on all levels in an organisation. 

Cybersecurity has become a large and multi-disciplinary domain. And we have been building this domain for a few decades now. Although a lot of progress has been made, I believe the cybersecurity field starts showing signs of aging. I will describe three drivers that are, in my view, the main contributors to the aging of cybersecurity: complexity of the cyber system, the incentives of the IT- and security industry, and the fundamental imperfection of security decision making due to uncertainty.

\section{complexity}
\begin{figure}[H]
\includegraphics[width=0.5\textwidth]{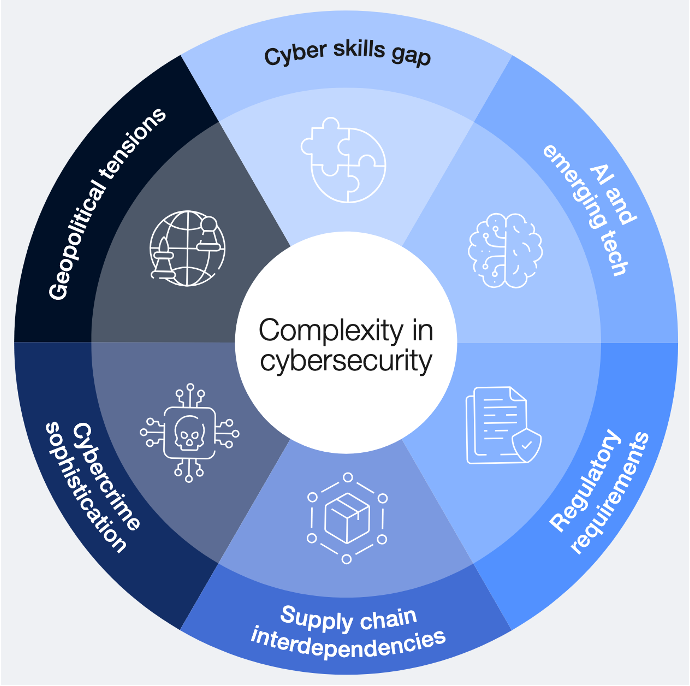}
\caption{\cite{wef}}\label{fig_wef}
\end{figure}

Digitalisation has changed the world and is impacting all organisations. The rise of new technologies like artificial intelligence and increased communication speeds of mobile 5G networks, our society is becoming more digital at increasing speed. Security of the digital domain has been assumed for a long time. Users of software assumed that their data would be safe and that the services would be available when they needed them. But nowadays the experience of users is very different. Next to the incidents I mentioned at the beginning of my lecture, there are examples in the newspapers every day. Examples of data being stolen, companies being locked up by ransomware or services being unavailable due to DDoS attacks (distributes denial of service). The complexity of the digital world has grown significantly and so has the field of cyber security (figure \ref{fig_wef}).

The economic effects of these incidents can be significant, and companies can even go bankrupt as a consequence. In 2024, the ECB shared in a presentation (\cite{ecb}), results of an Oliver Wyman report (\cite{OW}) that lists cyber risk as the risk of highest concern for the banking sector. Where banks used to worry about credit risks, they now worry about cyber risks.

Why is this? I think this is because more services are delivered through software. And that software is increasingly delivered to an organisation by a multitude of different other companies, like cloud providers. Therefore, organisations are becoming increasingly entangled with each other via long supply chains. This entangled eco-system has nodes of companies that are suppliers to many other organisations, and these nodes are vulnerable and concentrations of risk. It is not surprising that these vulnerabilities exist, because the eco-system is not designed but it grows due to individual choices of companies. Security by design is therefore not applicable to eco-systems: there is no designer.
\\

In the beautiful book “Overcomplicated: Technology at the Limits of Comprehension” by Samuel Arbesman (\cite{arbesman}), he describes how systems become more complex and how strange effects can emerge. The book contains many examples including the following one. The 1040 tax laws in the USA have become so complex, that judges have ruled that people cannot be blamed for willful failure to file tax returns (see \cite{arbesman}, p. 40). Obviously the 1040 tax system has become vulnerable due to its own complexity.

Software products are not a priori secure; they contain many vulnerabilities. These vulnerabilities can be used by cyber criminals or they manifest themselves as unintended behaviour of the system what can lead to damage.
Every day new vulnerabilities are discovered and published. Bigger organisations can have hundreds of thousands of instances of vulnerabilities, and their security teams are spending significant amounts of time on patching them. Some vulnerabilities are not discovered by the software producer, but they are discovered by hackers. These vulnerabilities are not always published or patched, but they are traded between criminals to be exploited in their attacks. While it is certain that software is vulnerable, the extend of that vulnerability is therefore very uncertain. This uncertainty is used by an increasing cybersecurity industry.
\\

An example of vulnerable software that was used on a massive scale leading to societal impact is the software of Citrix. The software of Citrix is used by many organisations to allow remote access (for example from their homes) by their employees or third parties. On December 17th, 2019, Citrix published a vulnerability in their software that allowed unauthorised parties access to the organisations using Citrix (the Citrix support article about this issue has been removed, but one can find details on github: \cite{citrix}). Citrix also published mitigating measures that could be used to deal with the vulnerability. But on January 17th, 2020, the National Cyber Security Centre of the Netherlands (NCSC) advised companies to switch off their Citrix implementations to avoid misuse (\cite{ncsc}). Clearly, the only way to secure the Citrix system was to switch it off. This was therefore a very impactful advice to the Netherlands.

\begin{figure}[H]
\includegraphics[width=0.5\textwidth]{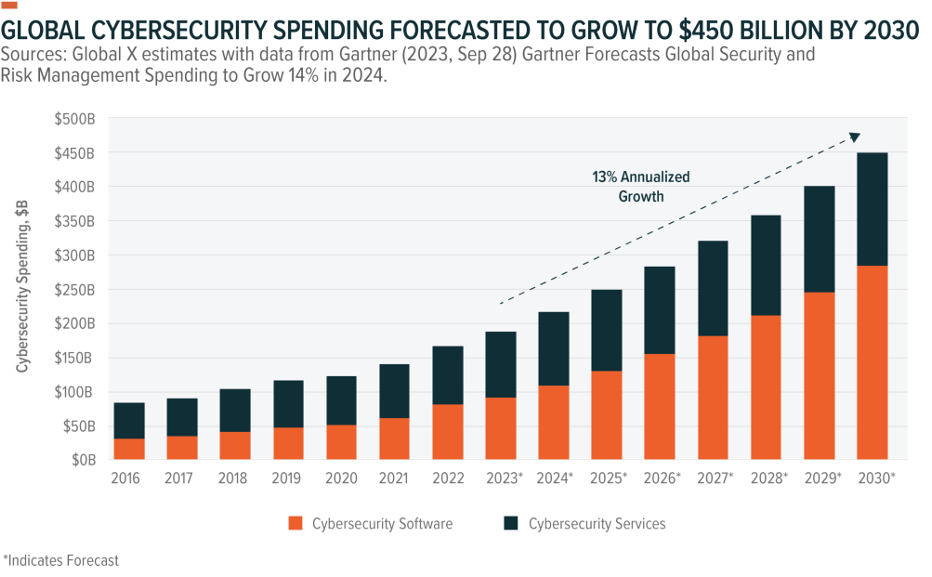}
\caption{}\label{fig_earnings}
\end{figure}

How is it possible that software that is so crucial for the correct functioning of an organisation, that is produced by a large global operating software company, with thousands of customers across the globe, is so vulnerable? The Dutch “Onderzoeksraad Voor Veiligheid” (OVV) investigated this and published the report “Kwetsbaar door Software” in 2021 (\cite{ovv}). I was involved in this research, and one of the things I learned was that cybersecurity has become a revenue model for the IT industry. In figure \ref{fig_earnings} one can see how much money is being made in that industry and how fast this is growing.

The OVV report investigated root causes of software being vulnerable and it concludes that the process of producing software inherently always introduces vulnerabilities. The main reasons are economical. Testing and repairing software costs money and time and commercial software producers make their own choices about when software is considered secure enough to sell. When vulnerabilities are discovered afterwards, patches will be provided, and end-users are expected to quickly apply them. 

This illuminates a fundamental problem: software end-users are depending on security decisions made by others and have no other choice than to accept the resulting uncertainty. I will talk about making security decisions under uncertainty later. 

On the other hand, software producers choose to allow the security of their products to depend on security decisions made by their end-users. Because of this, a very lucrative market for security products has emerged as we have seen. This market exploits the fact that software quality is low and software contains many vulnerabilities. And it exploits the fact that many organisations have become dependent on correct functioning of software. Indeed, security has become a revenue model of a fast growing security industry. This industry depends on software being weak and uncertainty about it being high.
\\

To be clear, I do believe we that security industry is useful and their services are essential. But that industry will not solve the root cause because the different incentives of involved parties are not aligned. That is why we see regulation stepping in.

\section{regulation}
The realisation that secure use of software has become necessary to build a viable digital society that works and in which people are protected, has led lawmakers, regulators and supervisors to establish multiple rules, laws and regulations. An example of a recent European law is Network and Information Security 2 (NIS2) directive (\cite{nis2}), which is now being transposed to local laws. For the financial industry the new Digital Operational Resiliency Act (DORA) is a lex specialis of NIS2 (\cite{dora}). 
I want to discuss a few specifics of these cybersecurity regulations. 

In the first place, these regulations are about cyber resilience, which is a broader notion than cybersecurity. 

Secondly, it recognizes that organisations are not isolated but operate in eco-systems consisting of partners and suppliers. We saw this in the Citrix incident as well. 

Lastly, these regulations lift the cybersecurity topic to a boardroom topic.
\\

How do these regulations do that? To increase cyber resilience, they make cybersecurity testing mandatory, through so-called hacker tests (“threat led penetration tests”). These and other tests help reduce uncertainty about the resiliency in organisations by discovering vulnerabilities in their systems and procedures. And by mitigating the test findings before malicious actors misuse them or before accidents happen.

The impact of dependencies on third parties means that parties in supply chains need to exchange information about their security posture, to allow all parties in the chain to make better security decisions. Both NIS2 and DORA contain clauses that make it mandatory to include such requirements in the legal contract underpinning the supply chain. 

Exchanging security posture information lowers uncertainty and is therefore useful, but without legal obligations this will not happen automatically.
With the intentions of these laws to reduce uncertainty and increase insights, they address one of the most important problems in the global cyber system. That is why I am very happy with these regulations: I love DORA.

The NIS2 and DORA regulations also contain clauses that make executive leaders in an organisation personally liable for blameworthy failing security setups. Hence the lawmaker is forcing the cybersecurity topic to be on the agenda of the executive board. Apparantly, this required the force of law.
This brings me to the central theme of my lecture: how are humans supposed to make good security decisions?

\section{uncertainty and risk}
We have seen that software contains many vulnerabilities and the information about the level of vulnerability is limited. Also, the complexity of software and systems is increasing via an accumulation of cloud services, suppliers and entanglement with other systems and organisations. As we saw, over time this has led to a massive increase in number of controls and security measures (NIST CSF framework grew from 400 to 1200 controls). Even if all these control implementations are regularly tested on operational effectiveness, their actual contribution to risk reduction is hard to assess and uncertain. So, there is limited information about vulnerabilities, limited information about threats, limited information about actual effectiveness of measures and limited information about cyber incidents that happened, their frequency and their impacts. But this type of information is necessary to calculate risks. For example, via the well-known formula 
\[
R = P \times I,
\]
meaning risk (R) equals the probability P times the impact I. As we just saw, both probability and impact can only be determined with a high level of uncertainty. Note that I distinguish between probability and uncertainty. 

\begin{figure}[H]
\includegraphics[width=0.5\textwidth]{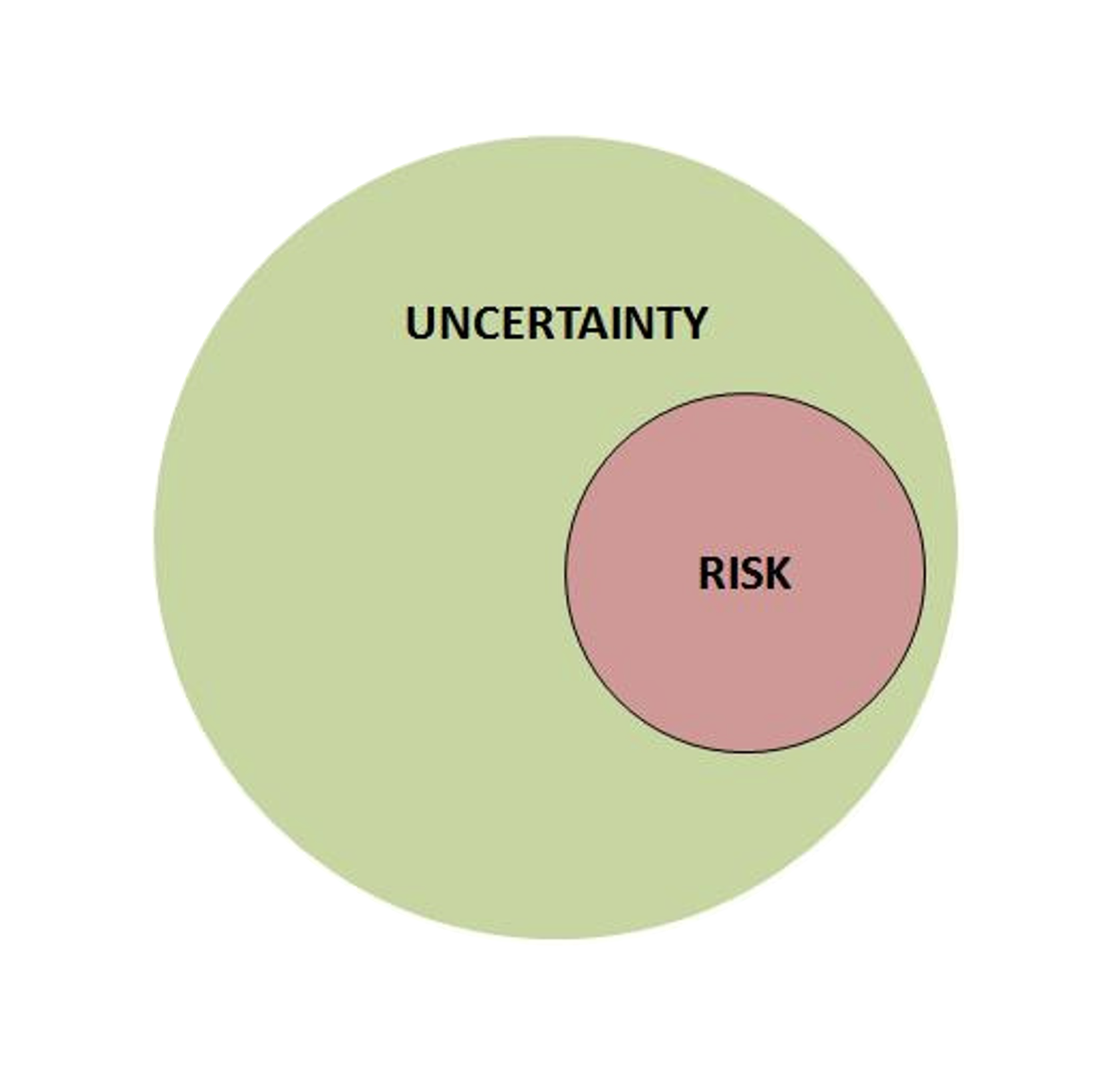}
\caption{uncertainty and risk}\label{a}
\end{figure}

Uncertainty means that we have so little (reliable) information that we cannot even guess the probability or impact of events happening in the future. So risk management is a subset of uncertainty management (see figure \ref{a}). Yet, through lack of better means, we do see that cyber risk management has become a common way to make security decisions.
Risk management does indeed sound promising. If we are not able to prevent security incidents to happen, at least we can try to calculate the risks and then implement measures that lower those risks. An entire industry has emerged that markets “cyber risk quantification” products. Given the discussions above, I believe that risk management is not precise enough to be able support good enough security decision making. Risk management works best in domains with high predictability. But as we have seen, the cyber security domain has low predictability, on longer time frames and we can no longer justify restricting attention only to the (assumed) predictable part (see also \cite{dekker-trans}). 

People are used to apply risk management to support decision making, and I expect it will continue to be used in the cyber domain as well.

My research questions are specifically aimed at understanding long term resiliency of an unpredictable and aging cyber security system by managing uncertainty.

\section{decidability and calculemus}
To understand why uncertainty is a problem for decision making, I will take a short detour to the philosophical root of decidability and calculability.
Cybersecurity originates from the technical domain of computers, networks and software. Perhaps this is the reason that cybersecurity is still considered a pure technical, precise and measurable topic. It leads to the tendency to express security in numerical values from which one can simply calculate security decisions.

This optimism is not a recent phenomenon. Gottfried Wilhelm von Leibniz was a German philosopher and mathematician who lived from 1646 until 1716. One of his famous quotes was “calculemus!”, which means: “let us calculate!”. He was convinced that human reasoning can be reduced to calculation. This quote appears in several of his works (for example in \cite{leibniz}). He thought that any decision to be made, can be translated into a calculation. He therefore proposed that differences of opinion are not necessary, as one can simply calculate what is true or what is the best decision.

This optimism was felt in the mathematics until the early 20th century. In those days, mathematicians tried to prove that every true mathematical expression can be deduced from a small set of axioms, by simple logical steps. The most prominent mathematicians that worked on this were Bertrand Russel and Alfred North Whitehead. They wrote their famous book “principia mathematica” (\cite{russell}) in which they formalised the arithmetic of natural numbers. But the optimism that could be generalised to all of mathematics was destroyed in 1931. 

In that year, the mathematician Kurt Gödel proved his famous incompleteness theorem. This theorem states that any consistent formal system, of sufficient strength or complexity\footnote{This is an imprecise formulation. The theorem assumes the formal system is capable of expressing propositions about the natural numbers.}, contains true statements that cannot be proved in that system. If even mathematics is not decidable or calculable, how can one hope to reduce human reasoning to calculations? 

I think that cybersecurity has reached a level of complexity and uncertainty that many security decisions are no longer calculable. I think that cyber risk management is not more than a rough approximation of security, that can work on small timescales and local well defined and isolated environments, but not beyond those, or in general.

\section{unfalsifiability and waste}
Can systems be secure and in use? In 2015, the researcher Cormac Herley published an elegant paper (\cite{herley}) in the Proceedings of the National Academy of Sciences in the USA. In that paper he shows that a claim that states that a system is not secure, cannot be falsified. He showed that the set $Y$ is undecidable, where $Y$ is defined as the set of system statuses $x$ of a system $X$, with the property

\[
x \in 
\begin{cases}
\textbf{Y} \text{ if bad outcomes will be avoided,} \\
\textbf{$\overline{\textbf{Y}}$} \text{ otherwise.}
\end{cases}
\]

This means that one cannot prove that a system is secure. This paper made me think, and I think it can have far reaching consequences for the study of longer term security decision making. 

Herley’s result has as consequence, that when someone proposes to remove a security control, the claim that the remaining cyber security control set has become less secure, cannot be falsified. Because it is easy to convince decision makers that adding a control increases security, there is a fundamental asymmetry in security: decisions to add a control are easy, decisions to remove a control are hard. 

We have seen that the information available for security decision making contains many uncertainties. And now we see that, regardless the quality of information, security decision making can never be perfect.
\\

What does all this mean? A lot has changed since the time that cyberspace was a simple computer network with a few computers connected by telephone lines, in which the Reaper and the Creeper were chasing each other. The global cyber security system has become very large, fast and complex. With the rise of artificial intelligence (AI) that speed and complexity will only increase. Even though AI is often touted as the solution to any problem, in this case AI will make the problem bigger. Because AI will make complexity cheaper, the law of stretched systems (that states that any system will consume any new capability completely until it reached full capacity again) means that complexity will only increase (this law is formulated by Hirschhorn in 1997, and cited in \cite{stretched}).

Executives that are personally liable, will continue to have to deal with high levels of uncertainty in cybersecurity decision making, and they will rather be safe than sorry. As we have seen, this means there is a natural selection pressure on the cybersecurity system to add controls, while unfalsifiability of security claims adds friction to removal of controls.
This mechanism will lead to an accumulation of waste in the system: controls that are no longer necessary and could be replaced or removed but will stay. 

From the theory of resilience engineering, we know that a system that accumulates waste, becomes less resilient (see for example \cite{re}).

\section{cyber senescence}
In the beautiful book “Other minds: the octopus and the evolution of intelligent life” (\cite{minds}), Peter Godfrey-Smith describes the life of an octopus. This book inspired me because I read it as a metaphor of cyber security. 

An octopus is a highly advanced creature, with a complex body that can communicate by changing colours in the skin. Experiments have shown that octopuses have a conscience, can play, use tools, and learn from observation. Clearly, a lot of energy is invested in creating this body. And yet, an octopus has a very short life. After about 3 or 4 years, the body literally falls apart: eyes fall out, arms fall off and the octopus literally breaks into pieces and dies.

How is it possible that so much energy is spent to build such a complex creature as an octopus, only to let it age to quickly? It appears that the DNA of an octopus contains a lot of waste, that only expresses itself after a few years. The selection pressure during the evolution of the octopus was maximally focused on the short term. An octopus lives in challenging environments, and apparently it was more useful to quickly develop than to grow old. The long-term negative effects of genes were less important than the short-term benefits. Over many centuries, a lot of waste accumulated in the DNA of the octopus that leads to the swift aging after the reproductive age, when all the negative longer term effects manifest themselves.

\begin{figure}[H]
\includegraphics[width=0.5\textwidth]{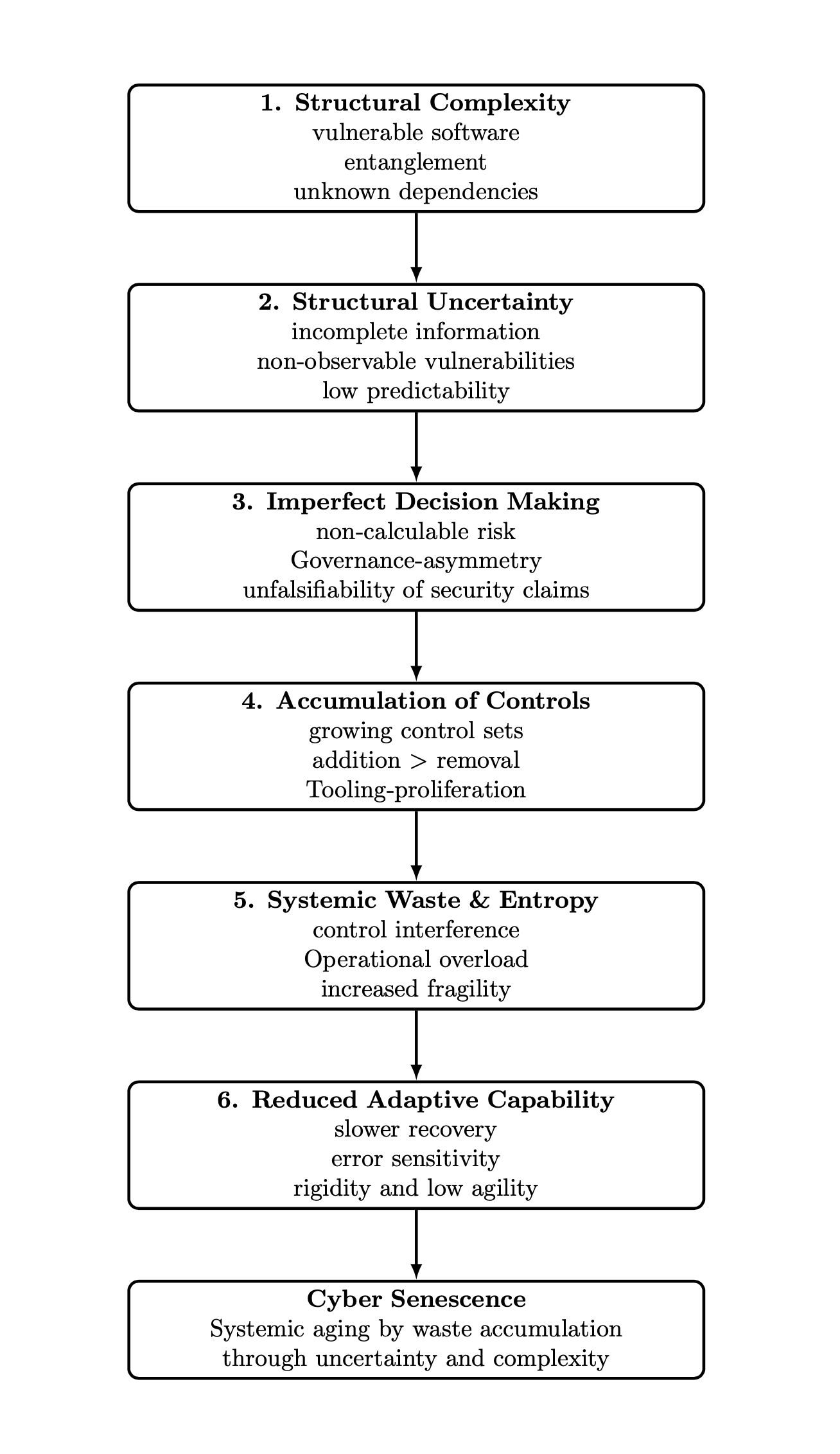}
\caption{the mechanism of cyber senescence}\label{fig_flow}
\end{figure}

How is life of an octopus a metaphor of cyber security? Cyber security has become a revenue model for the IT industry. Uncertainty about the real risks and the effectiveness of controls, combined with risk averseness of executives who start to feel their personal liability leads to high selection pressure to quickly patch vulnerabilities for every newly discovered threat. The agency problems that arise in security decision making and the biases that play into this are subject of recent study, see for example \cite{elif} and \cite{barre}. The unfalsifiability of security claims results in only adding controls regardless their long-term downsides. This leads to an accumulation of waste in the cyber security system. In other words: the cyber security system evolves in an eco-system that looks a lot like the environment in which an octopus evolved and in which all incentives point in the wrong direction.

Many years have passed since the Reaper and the Creeper. Our defence systems are starting to age. Because of the fundamental imperfection of security decisions, mistakes and waste accumulate and complexity of cyber defence itself will increase. This is already starting to manifest itself. The complexity of cloud computing has become so high, that cloud service providers now claim that one can only properly secure it by using the very newest technology like asset-graphs and algebras driven by AI (see for example \cite{graph}, a technology launched in 2025 to protect azure tenants, years after the launch of the azure cloud). This technology is not even ready yet, so how can we even start with security by design?

Sometimes, issues have become so large, that systems can only be protected by switching them off, like in the Citrix case in 2021. We are approaching the situation that we can no longer secure our digital world because both of the complexity of that world and because of the complexity of the security measures themselves.
\\

I propose that if we continue this way, the accumulation of waste in our security control frameworks will lead to decay. We are watching our cybersecurity systems aging, a new phenomenon that I call “cyber senescence”. 

\section{a cyber senescence research agenda}
How can we organise selection pressure on the system that can lead to sustainable security in a uncertain environment? I propose we need to do three things, and all three are part of the research of my chair’s mission.

\end{multicols}

\begin{figure}[H]

\includegraphics[width=\textwidth]{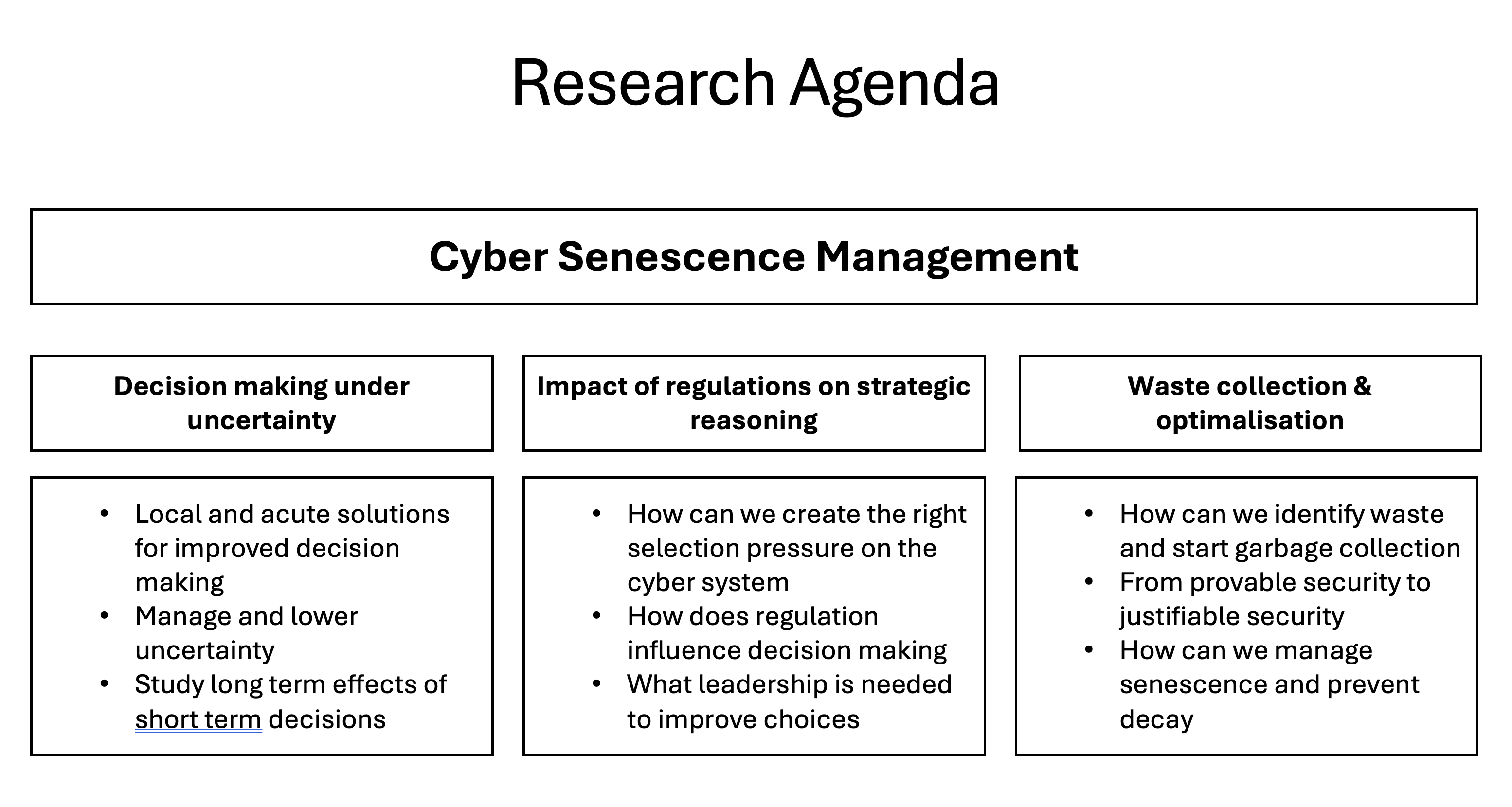}
\caption{research agenda}\label{fig_agenda}
\end{figure}

\begin{multicols}{2}

Firstly, we need to look for ways to make better local security decisions under uncertainty. This means we need to become better at detection of incidents, for example by applying AI to analyse the increasing volumes of telemetry, and make better threat analyses to reduce uncertainty and help choose controls. A lot of research is focussing on these problems and recently Lampis Alevizos and myself have published papers about this (\cite{dekker} and \cite{alevizos}).

Secondly, we need to find regulatory ways to create selection pressure on the global system that increases resilience. For this, it is important to understand how regulation, regulatory uncertainty and standards impact strategic choices in organisation. Eugenie Coche, a very talented PhD student at University of Amsterdam (supervised by Ans Kolk and myself) has already done important work in this field (\cite{coche}).

Thirdly and most importantly, we must learn how to manage cyber senescence. How can we cope with the accumulation of waste? How can we recognize waste and start garbage collection? How can we justify control removal and move from provable security to justifiable security? And if we cannot remove waste, how can we create new value from it? I believe a new, holistic and systemic analysis needs to be made of the entire cyber security system. How does this dynamic system evolve and how can we create selection pressure that avoids or at least manages cyber senescence?

All three topics require research as depicted in figure \ref{fig_agenda}, but it also requires education. It requires cyber security professionals that can make decisions or know how to effectively influence decision makers that lead to a resilient digital future. For this reason, I have set up the new Cybersecurity Leadership Academy at the Amsterdam Business School of the University of Amsterdam. I believe that the unique curriculum of this academy will help build a security workforce that combines deep understanding of the cyber security with knowledge about how decisions are made. It is this combination that will be needed for security professionals to be effective at influencing strategic decision making and managing cyber senescence.
\\

I will close. Cyberspace has become very complex, and it continues to develop through economic and technical choices by all parties that are part of it. Because of uncertainty, low predictability and imperfection of security decisions, wast accumulates in security controls frameworks which leads to cyber senescence: the decay of our digital world. Just like that beautiful octopus that falls apart after a short life.

My research and education will focus on developing insights into the mechanisms of cyber senescence and ways to deal with that.
This is my mission: to work on a digital future that is sustainable and in which humans are safe, free and happy.

\section{acknowledgements}
I would now like to express my thanks. I want to start with thanking the board of directors of the University of Amsterdam, the dean of the faculty and the “stichting tot bevordering van onderwijs en onderzoek in economie en bedrijfskunde”. In particular I want to thank the members of the curatorium of the chair Business \& Cybersecurity: prof. Marc Salomon, prof. Ilker Birbil and prof. Ans Kolk.

I want to thank Johan van Hall. A few years ago, while he was member of the executive board of ABN AMRO Bank N.V., he created the opportunity for me to have a part-time academic role next to my role as Chief Information Security Officer at that bank. Without his help, I would not stand here.

I want to thank Ans Kolk. Her support, advice and guidance since my start at the business school, were indispensable for me.  I want to thank Ans for her energy, insights and help she gave me to develop myself as a scholar.

Next I want to thank Marc Salomon for granting me a visiting position that provided me with the time and access I needed to build an academic CV. 

The cyber security field in The Netherlands knows many strong leaders and I want to thank one in particular: prof. Bibi van den Berg. My cooperation with Bibi started a few years ago at Centraal Bureau Statistiek (CBS) as members of their "Raad van Advies" (advisory board). From there, it developed into working together on publications and lectures. I want to thank Bibi for her inspiration, her help in my academic endeavours and for the pure fun of science she radiates.

I can only do the part-time role as professor by special appointment, because of the fantastic and strong security team in ABN AMRO Bank: thank you Coen, Rob, Sabine, Michael, Davina and Viola. I want to express my deep gratitude to Carsten Bittner and the members of the executive board of ABN AMRO Bank for their support and trust. Thank you.

All these people have played an important role. But I will now thank a few people in my life that have played an even larger role.
I want to thank my parents Bram and Trudy. Both are no longer alive. They have raised me in great freedom and always encouraged me to think independently.

Thank you, Esther, my sister, who is always there for me, to listen to me and to support me.

Thank you, Job, my twin brother I always look up to and with whom I will always continue our scientific debates. 28 years ago, we stood together in the aula of the University Utrecht, during your defence of your PhD thesis. And two weeks later we both stood here, in the aula of University of Amsterdam, during my defence of my PhD thesis (\cite{dekker-thesis}). Then we wore tailcoats and look at us now: both wearing toga!

I want to thank my daughter Annika and son Quinten: I am so happy that you are again part of my life.

Lastly, I want to thank my wife, Sophie. Her inspiration and slight nudges gave me the confidence to start developing and writing down my ideas. She is my muse. Thank you Sophie, my dearest.
There is so much to be thankful for in my life. But I will stop here. 
I have spoken.
\\

Ik heb gezegd.
\end{multicols}

\bibliographystyle{plain}
\bibliography{bib}{}

\end{document}